\documentclass[twocolumn,showpacs,preprintnumbers,amsmath,amssymb]{revtex4}

\usepackage{graphicx}
\usepackage{dcolumn}
\usepackage{bm}
\usepackage{epsfig}

\begin{document}


\title{\boldmath 
Reconciling $J/\psi$ production at HERA, RHIC, Tevatron, and LHC with
NRQCD factorization at next-to-leading order}

\author{Mathias Butensch\"on}
\author{Bernd A. Kniehl}
\affiliation{{II.} Institut f\"ur Theoretische Physik, Universit\"at Hamburg,
Luruper Chaussee 149, 22761 Hamburg, Germany}

\date{\today}

\begin{abstract}
We calculate the cross section of inclusive direct $J/\psi$ hadroproduction at
next-to-leading order (NLO) within the factorization formalism of
nonrelativistic quantum chromodynamics (NRQCD), including the full relativistic
corrections due to the intermediate $^1\!S_0^{[8]}$, $^3\!S_1^{[8]}$, and
$^3\!P_J^{[8]}$ color-octet states.
We perform a combined fit of the color-octet (CO) long-distance matrix
elements to the transverse-momentum ($p_T$) distributions measured by CDF at
the Fermilab Tevatron and H1 at DESY HERA and demonstrate that they also
successfully describe the $p_T$ distributions from PHENIX at BNL RHIC and CMS
at the CERN LHC as well as the photon-proton c.m.\ energy and (with worse
agreement) the inelasticity distributions from H1.
This provides a first rigorous test of NRQCD factorization at NLO.
In all experiments, the CO processes are shown to be indispensable.
\end{abstract}

\maketitle

The factorization formalism of nonrelativistic QCD (NRQCD) \cite{Bodwin:1994jh}
provides a rigorous theoretical framework for the description of
heavy-quarkonium production and decay.
This implies a separation of process-dependent short-distance coefficients, to
be calculated perturbatively as expansions in the strong-coupling constant
$\alpha_s$, from supposedly universal long-distance matrix elements
(LDMEs), to be extracted from experiment.
The relative importance of the latter can be estimated by means of velocity
scaling rules; {\it i.e.}, the LDMEs are predicted to scale with a definite
power of the heavy-quark ($Q$) velocity $v$ in the limit $v\ll1$.
In this way, the theoretical predictions are organized as double expansions in
$\alpha_s$ and $v$.
A crucial feature of this formalism is that it takes into account the complete
structure of the $Q\overline{Q}$ Fock space, which is spanned by the states
$n={}^{2S+1}L_J^{[a]}$ with definite spin $S$, orbital angular momentum
$L$, total angular momentum $J$, and color multiplicity $a=1,8$.
In particular, this formalism predicts the existence of color-octet (CO)
processes in nature.
This means that $Q\overline{Q}$ pairs are produced at short distances in
CO states and subsequently evolve into physical, color-singlet (CS) quarkonia
by the nonperturbative emission of soft gluons.
In the limit $v\to0$, the traditional CS model (CSM) is recovered in the case
of $S$-wave quarkonia.
In the case of $J/\psi$ production, the CSM prediction is based just on the
$^3\!S_1^{[1]}$ CS state, while the leading relativistic corrections, of
relative order ${\cal O}(v^4)$, are built up by the $^1\!S_0^{[8]}$,
$^3\!S_1^{[8]}$, and $^3\!P_J^{[8]}$ ($J=0,1,2$) CO states.

The greatest success of NRQCD was that it was able to explain the $J/\psi$
hadroproduction yield at the Fermilab Tevatron \cite{Cho:1995vh}, while the
CSM prediction lies orders of magnitude below the data, even if the latter
is evaluated at NLO \cite{Campbell:2007ws,Gong:2008ft}.
Also in the case of $J/\psi$ photoproduction at DESY HERA, the CSM cross
section at NLO significantly falls short of the data
\cite{Kramer:1994zi,Butenschoen:2009zy}.
Complete NLO calculations in NRQCD were performed for inclusive $J/\psi$
production in two-photon collisions \cite{Klasen:2001cu}, $e^+e^-$ annihilation
\cite{Zhang:2009ym}, and direct photoproduction \cite{Butenschoen:2009zy}.
As for hadroproduction at NLO, the CO contributions due to intermediate
$^1\!S_0^{[8]}$ and $^3\!S_1^{[8]}$ states \cite{Gong:2008ft} were calculated
as well as the complete NLO corrections to $\chi_J$ production, including the
$^3\!S_1^{[8]}$ contribution \cite{Ma:2010vd}.

In order to convincingly establish the CO mechanism and the LDME universality,
it is an urgent task to complete the NLO description of $J/\psi$
hadroproduction by including the full CO contributions at NLO, which is
actually achieved in this Letter.
In fact, because of their high precision and their wide coverage and fine
binning in $p_T$, the Tevatron data on inclusive $J/\psi$ production have so
far been the major source of information on the CO LDMEs \cite{Kniehl:1998qy},
and the LHC data to come will be even more constraining.
Previous NLO analyses of $J/\psi$ hadroproduction 
\cite{Campbell:2007ws,Gong:2008ft} were lacking the
$^3\!P_J^{[8]}$ contributions, for which there is no reason to be
insignificant.
This technical bottleneck, which has prevented essential progress in the global
test of NRQCD factorization for the past fifteen years, is overcome here by
further improving and refining the calculational techniques developed in
Ref.~\cite{Butenschoen:2009zy}.

Invoking the factorization theorems of the QCD parton model and NRQCD
\cite{Bodwin:1994jh}, the inclusive $J/\psi$ hadroproduction cross section is
evaluated from
\begin{eqnarray}
d\sigma(AB\to J/\psi+X)
&=&\sum_{i,j,n} \int dxdy\, f_{i/A}(x)f_{j/B}(y)
\label{Overview.Cross}\\
&&{}\times\langle{\cal O}^{J/\psi}[n]\rangle
d\sigma(i j\to c\overline{c}[n]+X),
\nonumber
\end{eqnarray}
where $f_{i/A}(x)$ are the parton distribution functions
of hadron $A$,
$\langle{\cal O}^{J/\psi}[n]\rangle$ are the LDMEs, and
$d\sigma(i j\to c\overline{c}[n]+X)$ are the partonic cross sections.
Working in the fixed-flavor-number scheme, $i$ and $j$ run over the gluon $g$
and the light quarks $q=u,d,s$ and anti-quarks $\overline q$.
The counterpart of Eq.~(\ref{Overview.Cross}) for direct photoproduction
emerges by replacing $f_{i/A}(x)$ by the photon flux function $f_{\gamma/e}(x)$
and fixing $i=\gamma$.

We checked analytically that all appearing singularities cancel.
As for the ultraviolet singularities, we renormalize the charm-quark mass and
the wave functions of the external particles according to the on-shell scheme,
and the strong-coupling constant according to the modified minimal-subtraction
scheme.
The infrared (IR) singularities are canceled similarly as described in
Ref.~\cite{Butenschoen:2009zy}.
In particular, the $^3\!P_J^{[8]}$ short-distance cross sections produce two
new classes of soft singularities, named {\em soft~\#2} and {\em soft~\#3}
terms, on top of the {\em soft~\#1} terms familiar from the $S$-wave channels.
The soft~\#2 terms do not factorize to LO cross sections;
they cancel against the IR singularities of the virtual corrections left over
upon the usual cancellation against the soft~\#1 terms.
The soft~\#3 terms cancel against the IR singularities  from the radiative
corrections to the $\langle {\cal O}^{J/\psi}(^3\!S_1^{[1]}) \rangle$ and
$\langle {\cal O}^{J/\psi}(^3\!S_1^{[8]}) \rangle$ LDMEs.
\begin{figure*}
\begin{tabular}{ccc}
\includegraphics[width=5.8cm]{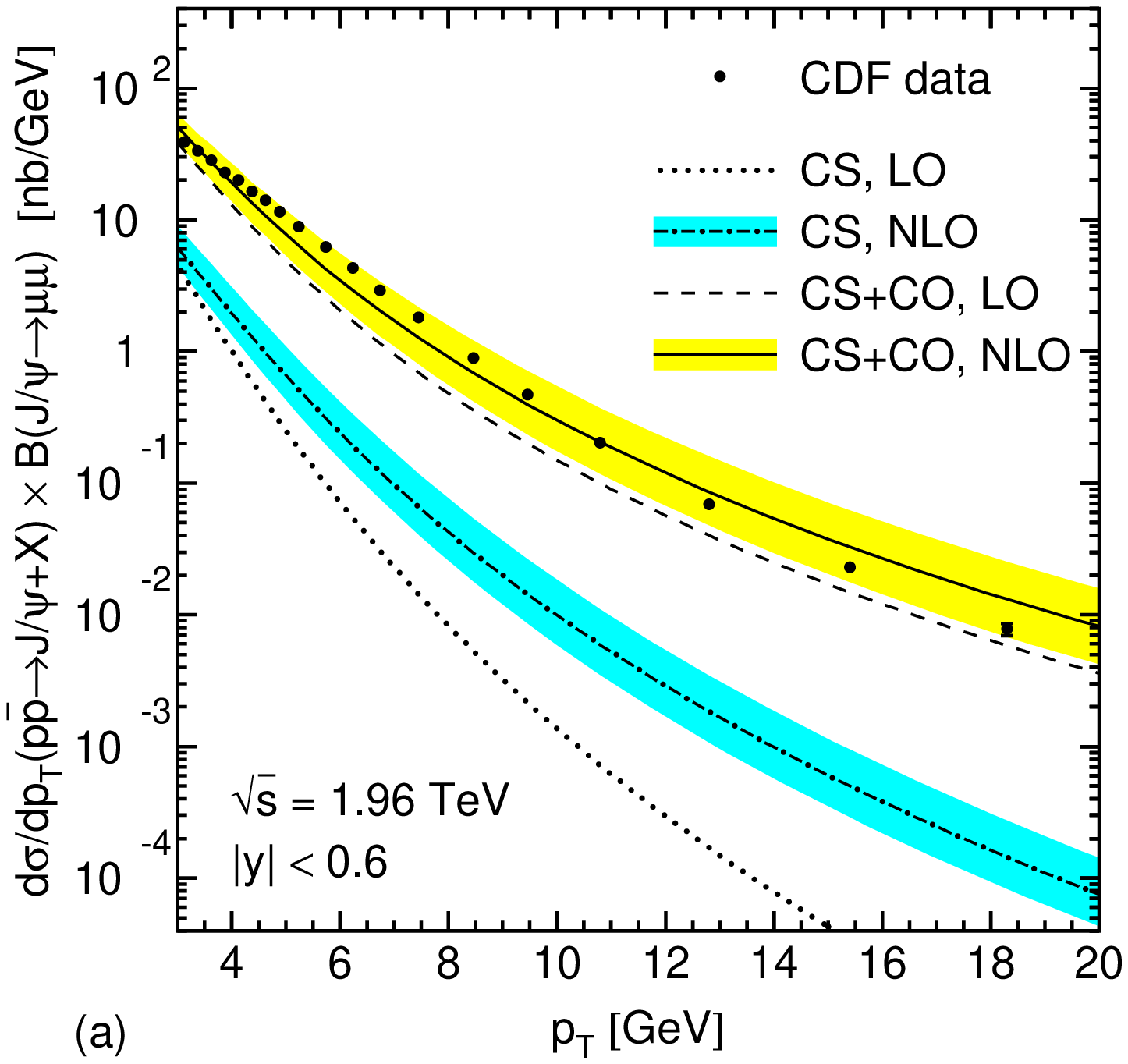}
&
\includegraphics[width=5.8cm]{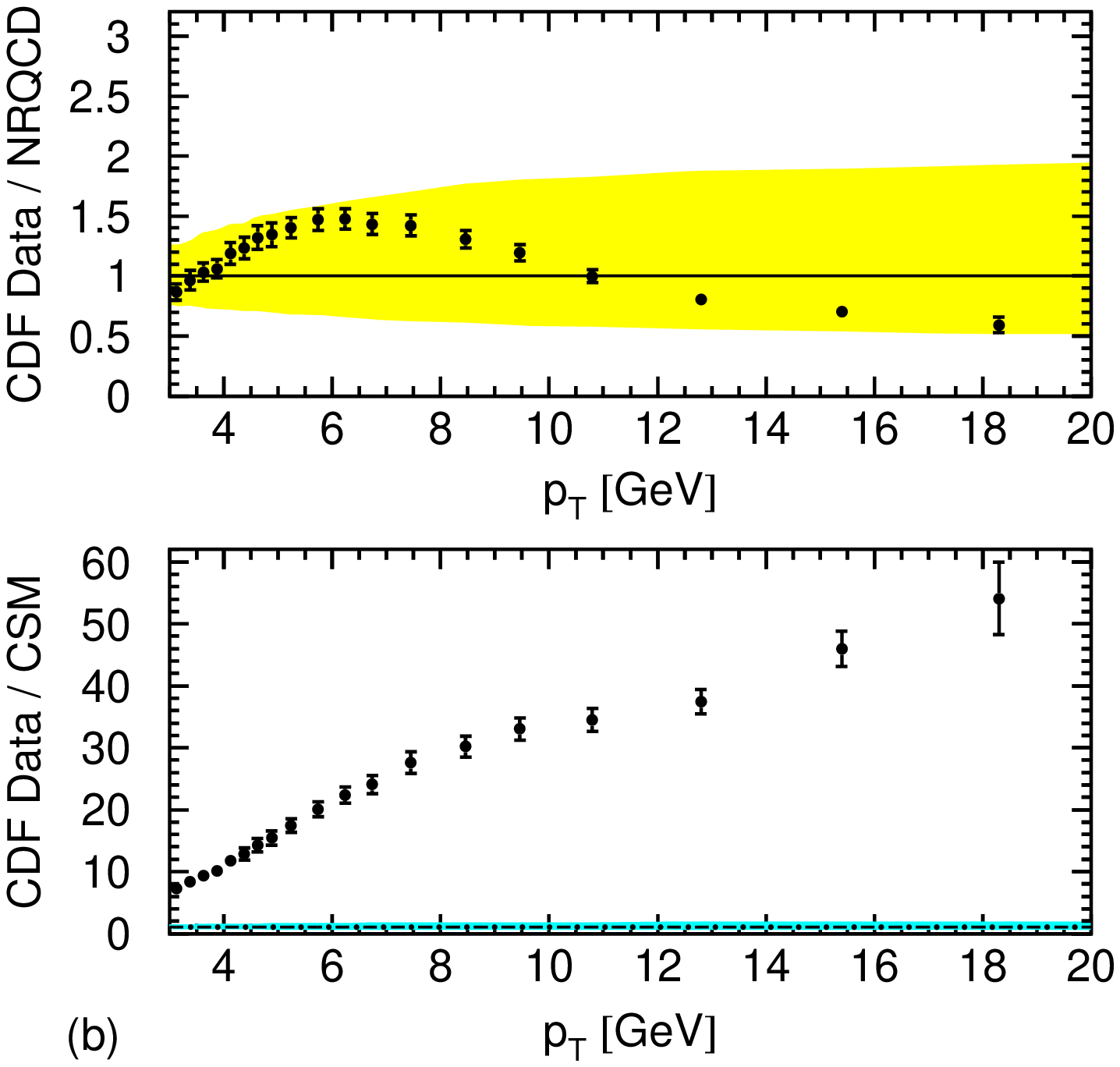}
&
\includegraphics[width=5.8cm]{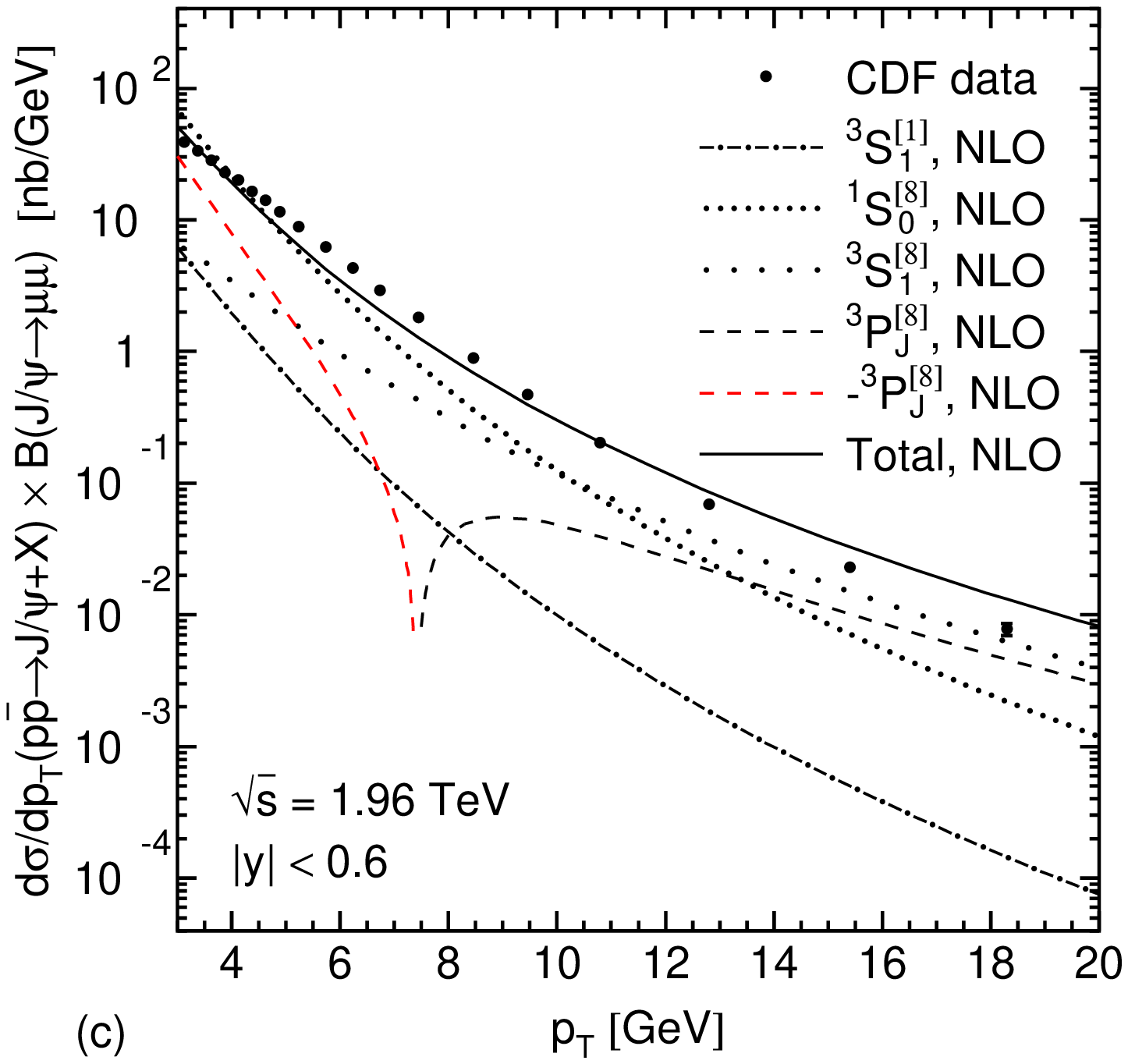}
\\
\includegraphics[width=5.8cm]{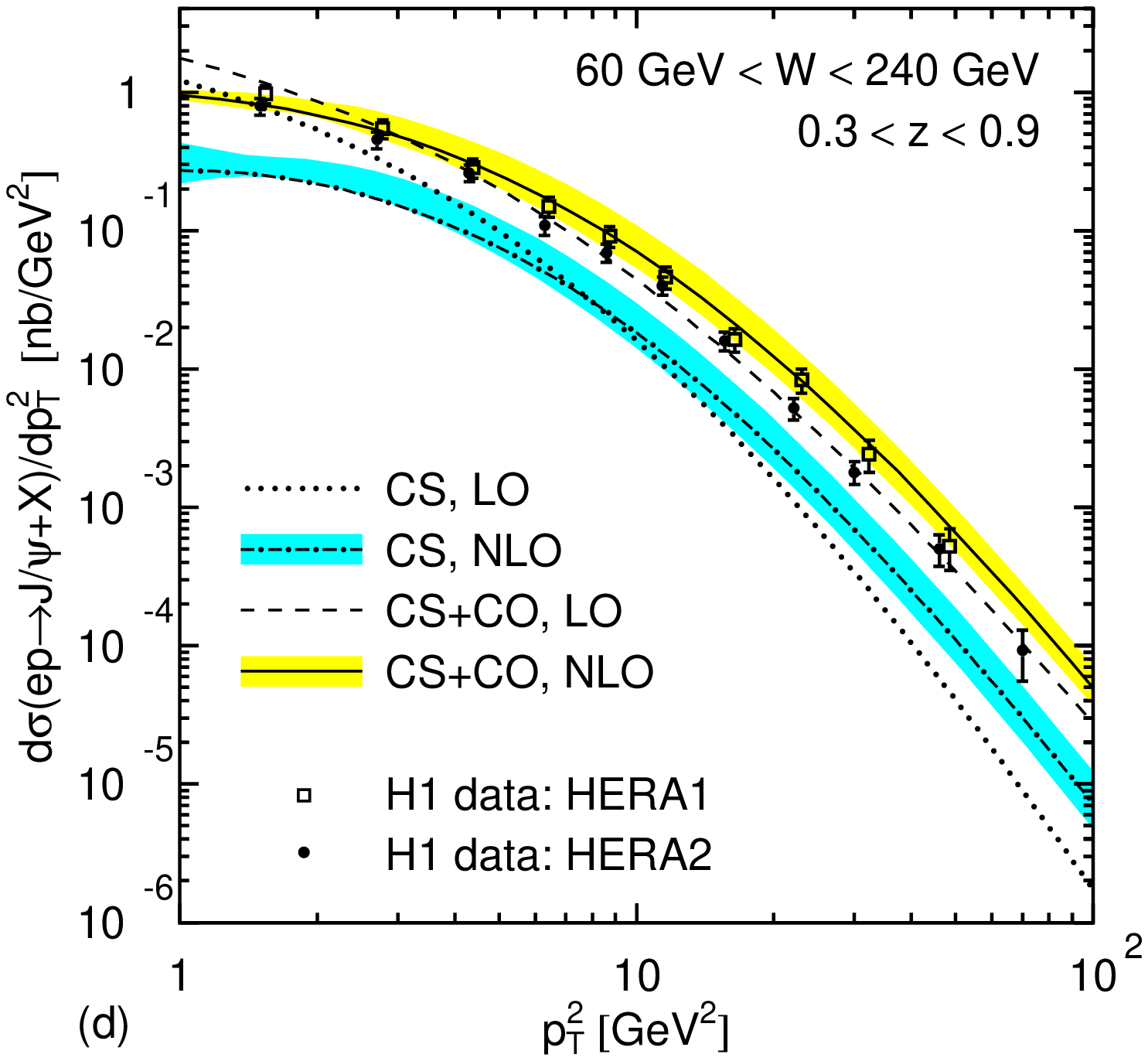}
&
\includegraphics[width=5.8cm]{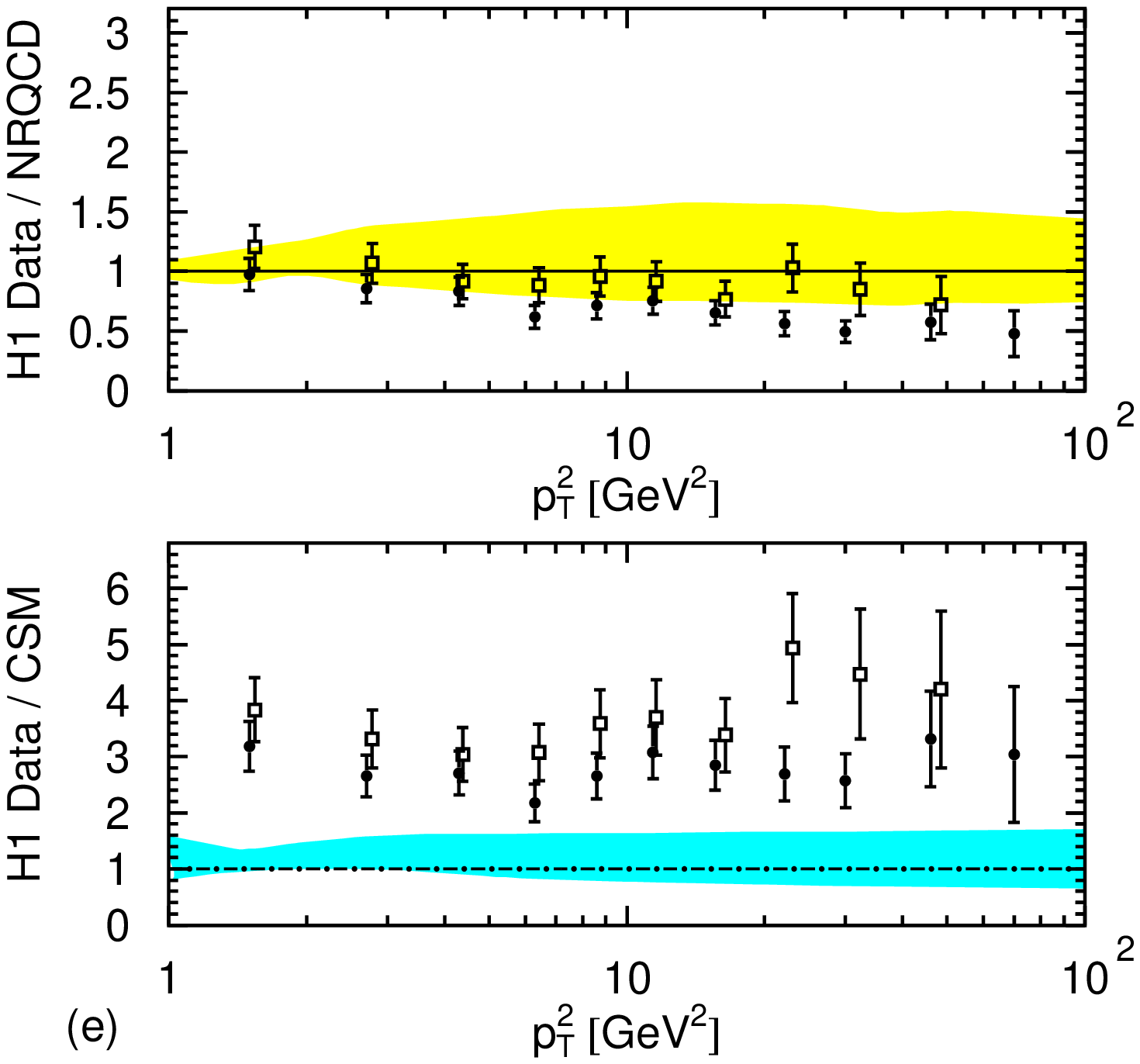}
&
\includegraphics[width=5.8cm]{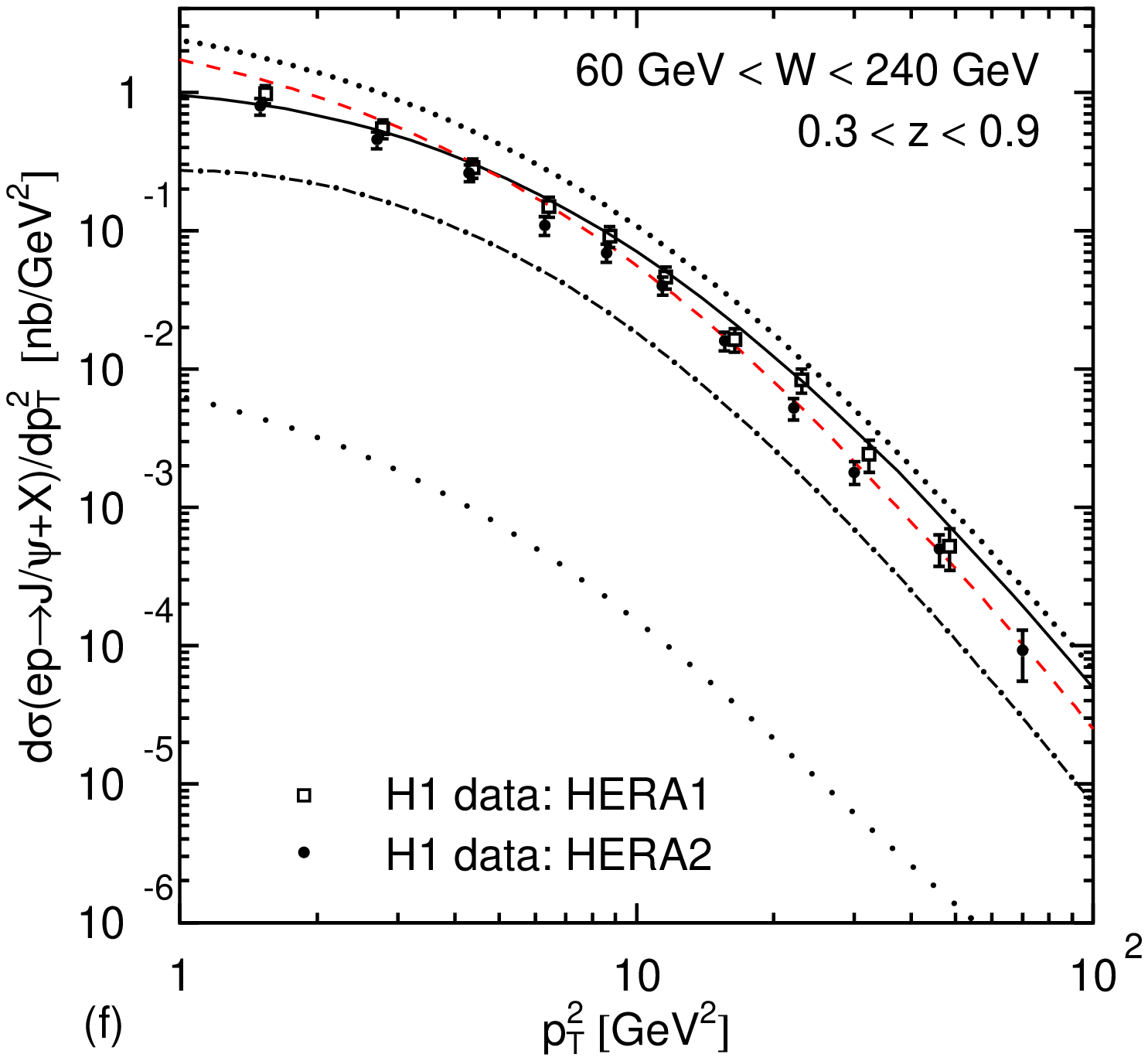}
\end{tabular}
\caption{\label{fig:fitgraphs}
NLO NRQCD predictions of $J/\psi$ hadro- and photoproduction resulting from the
fit compared to the CDF \cite{Acosta:2004yw} and H1
\cite{Adloff:2002ex,Aaron:2010gz} input data.
The coding of the lines in part (f) of the figure is the same as in part (c).
The seeming singularity of the $^3\!P_J^{[8]}$ contribution in part (c) is an
artifact of the logarithmic scale on the vertical axis..  
}
\end{figure*}

We now describe our theoretical input and the kinematic conditions for our
numerical analysis.
We set $m_c=1.5$~GeV, adopt the values of $m_e$, $\alpha$, and the branching
ratios  $B(J/\psi\to e^+e^-)$ and $B(J/\psi\to\mu^+\mu^-)$ from
Ref.~\cite{Nakamura:2010zzi}, and use the one-loop (two-loop) formula for
$\alpha_s^{(n_f)}(\mu)$, with $n_f=4$ active quark flavors, at LO (NLO).
As for the proton parton distribution functions,
we use set CTEQ6L1 (CTEQ6M) \cite{Pumplin:2002vw} at
LO (NLO), which comes with an asymptotic scale parameter of
$\Lambda_\mathrm{QCD}^{(4)}=215$~MeV (326~MeV).
We evaluate the photon flux function using Eq.~(5) of
Ref.~\cite{Kniehl:1996we} with the cut-off $Q_\mathrm{max}^2=2.5$~GeV$^2$
\cite{Aaron:2010gz} on the photon virtuality.
As for the CS LDME, we adopt the value
$\langle {\cal O}^{J/\psi}(^3\!S_1^{[1]}) \rangle = 1.32$~GeV$^3$ from
Ref.~\cite{Bodwin:2007fz}.
Our default choices for the renormalization, factorization, and NRQCD scales
are $\mu_r=\mu_f=m_T$ and $\mu_\Lambda=m_c$, respectively, where
$m_T=\sqrt{p_T^2+4m_c^2}$ is the $J/\psi$ transverse mass.

\begin{figure*}
\begin{tabular}{cc}
\includegraphics[width=5.8cm]{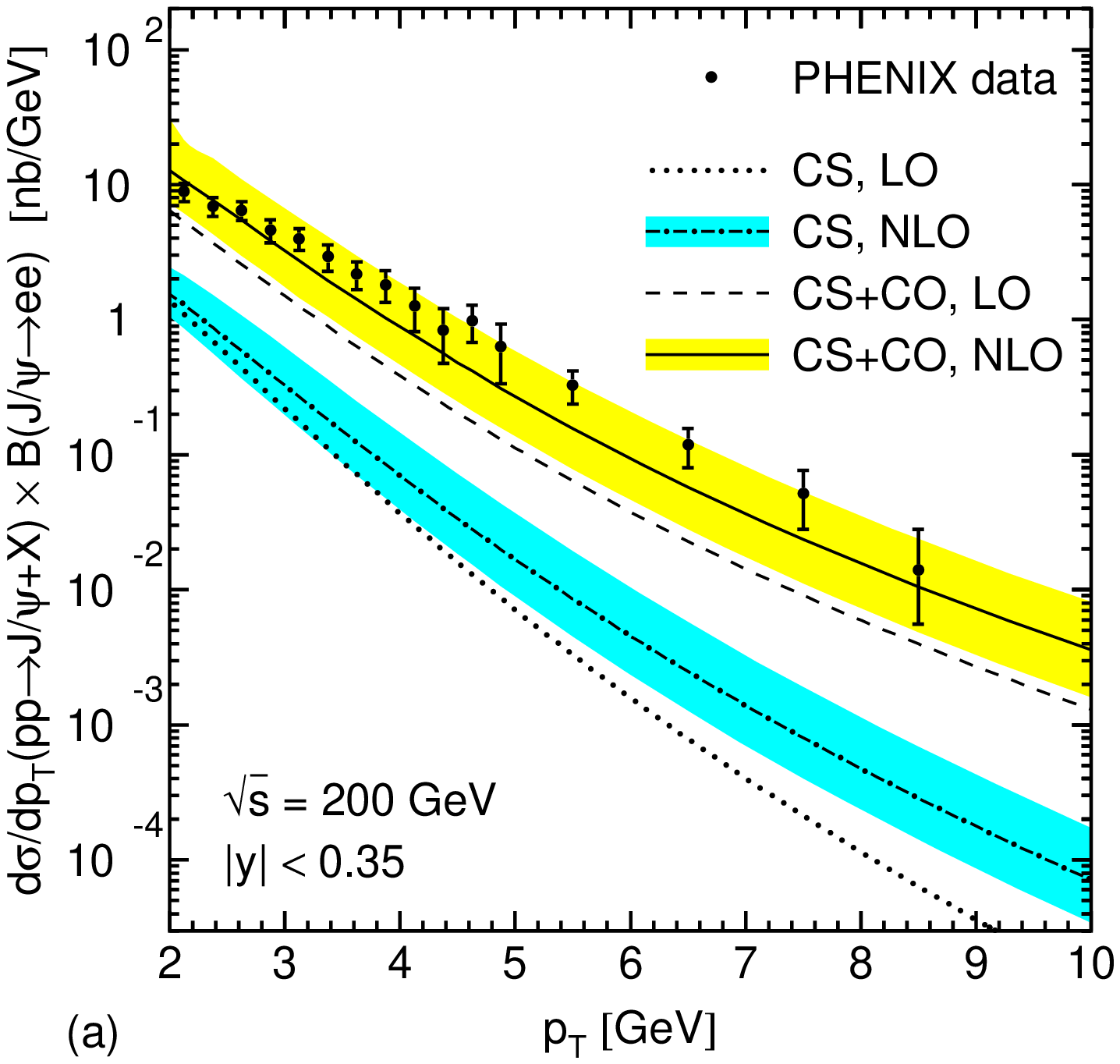}
&
\includegraphics[width=5.8cm]{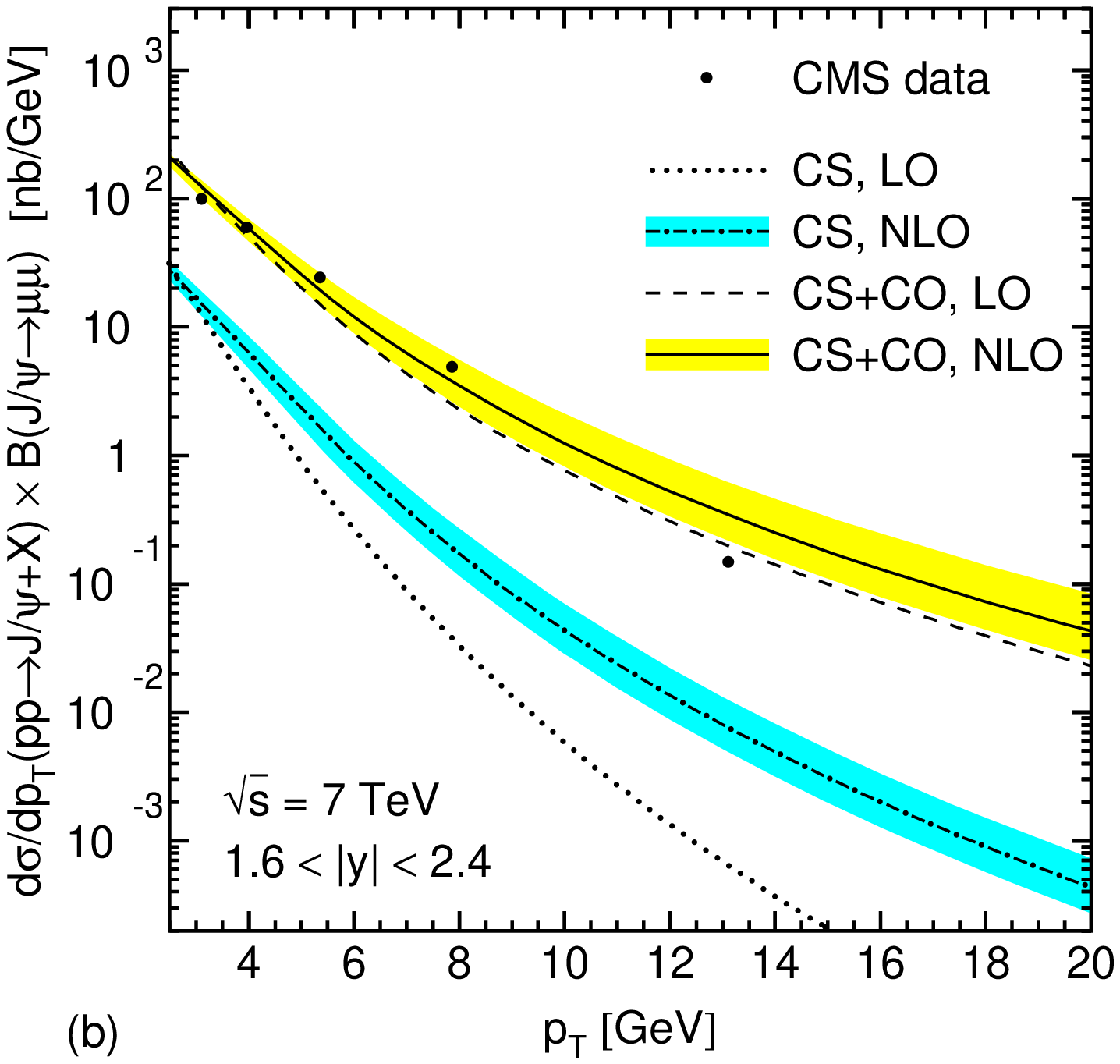} \\

\includegraphics[width=5.8cm]{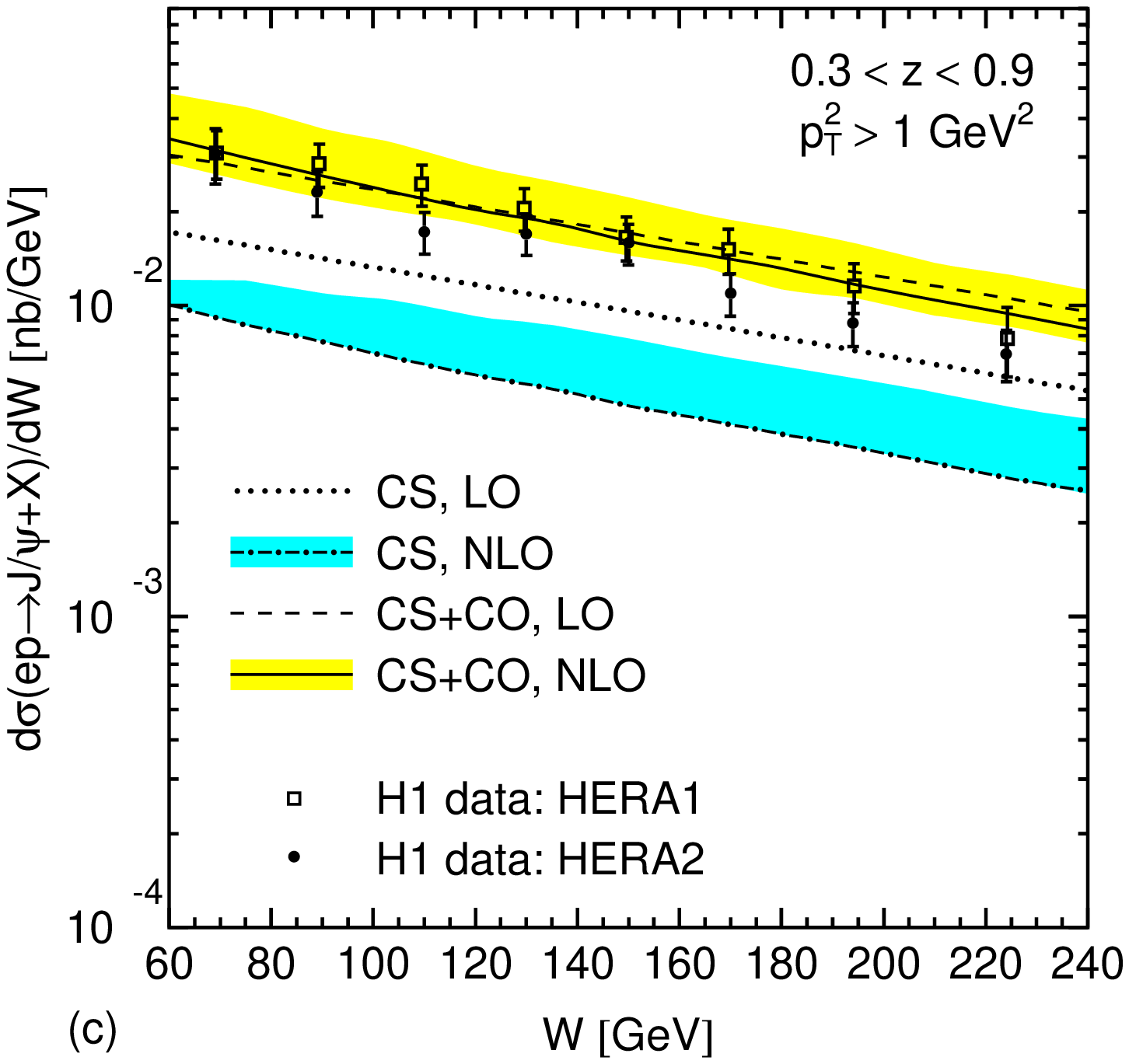}
&
\includegraphics[width=5.8cm]{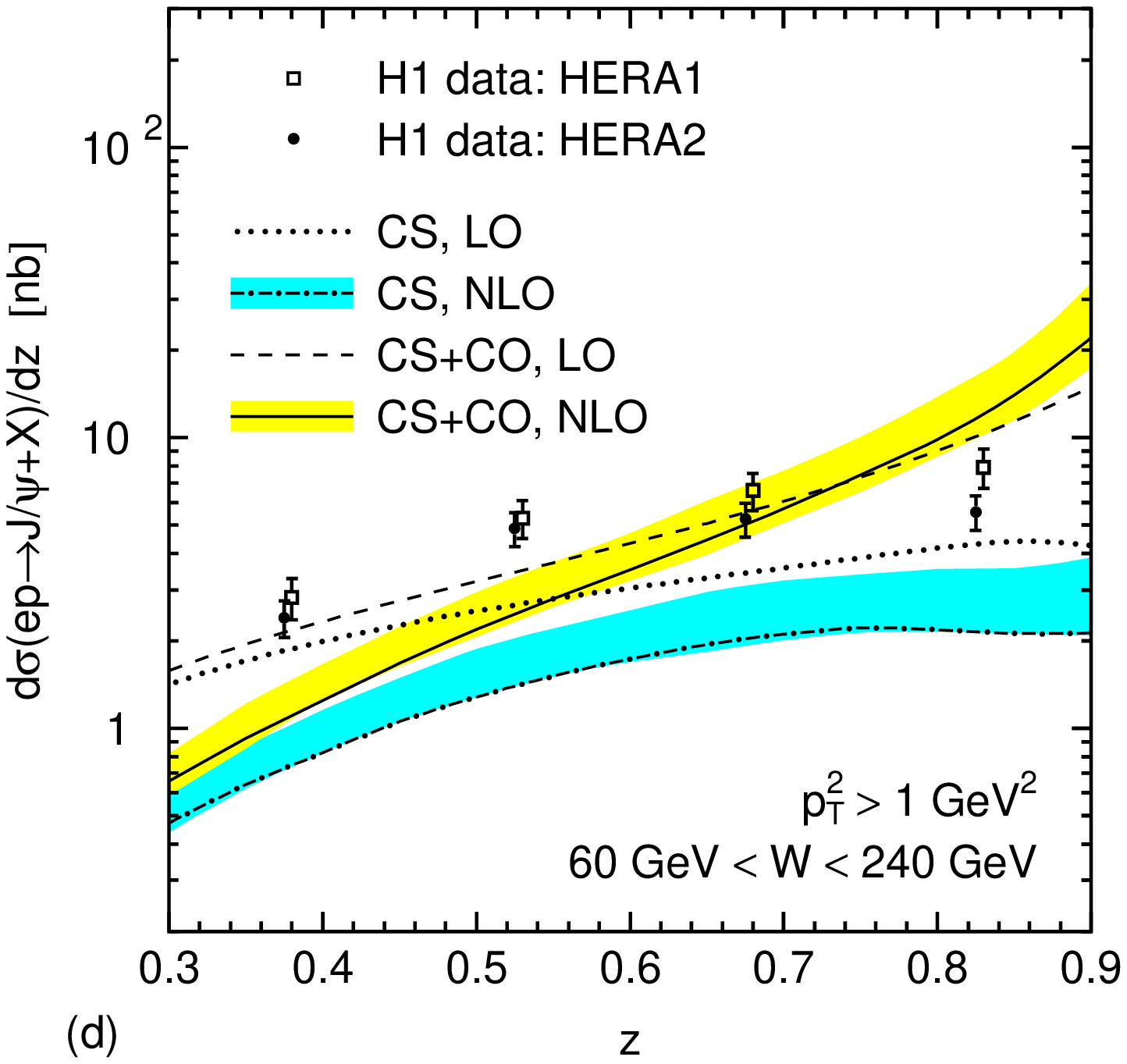}
\end{tabular}
\caption{\label{fig:other}
NLO NRQCD predictions of $J/\psi$ hadro- and photoproduction resulting from the
fit compared to PHENIX \cite{Adare:2009js}, CMS \cite{Collaboration:2010yr},
and H1 \cite{Adloff:2002ex,Aaron:2010gz} data not included in the fit.}
\end{figure*}

Our strategy for testing NRQCD factorization in $J/\psi$ production at NLO is
as follows.
We first perform a common fit of the CO LDMEs to the $p_T$ distributions
measured by CDF in hadroproduction at Tevatron Run II \cite{Acosta:2004yw} and
by H1 in photoproduction at HERA1 \cite{Adloff:2002ex} and HERA2
\cite{Aaron:2010gz} (see Table~\ref{tab:fit} and Fig.~\ref{fig:fitgraphs}).
We then compare the $p_T$ distributions measured by PHENIX at RHIC
\cite{Adare:2009js} and CMS at the LHC \cite{Collaboration:2010yr} as well as
the $W$ and $z$ distributions measured by H1 at HERA1 \cite{Adloff:2002ex} and
HERA2 \cite{Aaron:2010gz} with our respective NLO predictions based on these CO
LDMEs (see Fig.~\ref{fig:other}).

\begin{table}
\begin{tabular}{|c|c|}
\hline
$\langle {\cal O}^{J/\psi}(^1\!S_0^{[8]}) \rangle$ &
$(4.50\pm0.72)\times10^{-2}$~GeV$^3$ \\
$\langle {\cal O}^{J/\psi}(^3\!S_1^{[8]}) \rangle$ &
$(3.12\pm0.93)\times10^{-3}$~GeV$^3$ \\
$\langle {\cal O}^{J/\psi}(^3\!P_0^{[8]}) \rangle$ &
$(-1.21\pm0.35)\times10^{-2}$~GeV$^5$ \\
\hline
\end{tabular}
\caption{\label{tab:fit} NLO fit results for the $J/\psi$ CO LDMEs.}
\end{table}
The $p_T$ distribution of $J/\psi$ hadroproduction measured experimentally
flattens at $p_T<3$~GeV due to nonperturbative effects, a feature that cannot
be faithfully described by fixed-order perturbation theory.
We, therefore, exclude the CDF data points with $p_T<3$~GeV from our fit.
We checked that our fit results depend only feebly on the precise location of
this cut-off.
We also verified that exclusion of the H1 data points with $p_T<2.5$~GeV, which
might require power corrections neglected here, is inconsequential for our fit.
The fit results for the CO LDMEs corresponding to our default NLO NRQCD
predictions are collected in Table~\ref{tab:fit}.
In Figs.~\ref{fig:fitgraphs}(a) and (d), the latter (solid lines) are compared
with the CDF \cite{Acosta:2004yw} and H1 \cite{Adloff:2002ex,Aaron:2010gz}
data, respectively.
For comparison, also the default predictions at LO (dashed lines) as well as
those of the CSM at NLO (dot-dashed lines) and LO (dotted lines) are shown.
In order to visualize the size of the NLO corrections to the hard-scattering
cross sections, the LO predictions are evaluated with the same LDMEs.
The yellow and blue (shaded) bands indicate the theoretical errors on the
NLO NRQCD and CSM predictions, respectively, due to the lack of knowledge of
corrections beyond NLO, which are estimated by varying $\mu_r$, $\mu_f$, and
$\mu_\Lambda$ by a factor 2 up and down relative to their default values.
The $\mu_r$, $\mu_f$, and $\mu_\Lambda$ dependencies of $\alpha_s$, the 
parton distribution functions,
and the LDMEs, respectively, induced by the renormalization group are
canceled only partially, namely through the order of the calculation, by
linearly logarithmic terms appearing in the NLO corrections.
Data-over-theory representations of Figs.~\ref{fig:fitgraphs}(a) and (d) are
given in Figs.~\ref{fig:fitgraphs}(b) and (e), respectively.
In Figs.~\ref{fig:fitgraphs}(c) and (f), the default NLO NRQCD predictions of
Figs.~\ref{fig:fitgraphs}(a) and (d), respectively, are decomposed into their
$^3\!S_1^{[1]}$, $^1\!S_0^{[8]}$, $^3\!S_1^{[8]}$, and $^3\!P_J^{[8]}$
components.
We observe from Fig.~\ref{fig:fitgraphs}(c) that the $^3\!P_J^{[8]}$
short-distance cross section of hadroproduction (excluding the negative LDME)
receives sizable NLO corrections that even turn it negative at $p_T\agt7$~GeV.
This is, however, not problematic because a particular CO contribution
represents an unphysical quantity depending on the choices of renormalization
scheme and scale $\mu_\Lambda$ and is entitled to become negative as long as
the full cross section remains positive.
Such features are familiar, e.g., from inclusive heavy-hadron production at
NLO \cite{Kniehl:2004fy}.
In contrast to the situation at LO, the line shapes of the $^1\!S_0^{[8]}$ and
$^3\!P_J^{[8]}$ contributions significantly differ at NLO, so that
$\langle {\cal O}^{J/\psi}(^1\!S_0^{[8]}) \rangle$ and
$\langle {\cal O}^{J/\psi}(^3\!P_0^{[8]}) \rangle$ may now be fitted
independently (see Table~\ref{tab:fit}).
Besides that, the injection of HERA data into the fit also supports the
independent determination of $\langle {\cal O}^{J/\psi}(^1\!S_0^{[8]}) \rangle$
and $\langle {\cal O}^{J/\psi}(^3\!P_0^{[8]}) \rangle$.
Notice that $\langle {\cal O}^{J/\psi}(^3\!P_0^{[8]}) \rangle$ comes out
negative, which is not problematic by the same token as above.
In compliance with the velocity scaling rules of NRQCD \cite{Bodwin:1994jh},
the CO LDMEs in Table~\ref{tab:fit} are approximately of order ${\cal O}(v^4)$
relative to $\langle {\cal O}^{J/\psi}(^3\!S_1^{[1]}) \rangle$.
We read off from Fig.~\ref{fig:fitgraphs}(a) and (d) that the NLO correction
($K$) factors for hadro- and photoproduction range from 1.30 to 2.28 and from
0.54 to 1.80, respectively, in the $p_T$ intervals considered.

We observe from Fig.~\ref{fig:other} that our NLO NRQCD predictions nicely
describe the $p_T$ distributions from PHENIX \cite{Adare:2009js} (a) and CMS
\cite{Collaboration:2010yr} (b) as well as the $W$ distributions from H1
\cite{Adloff:2002ex,Aaron:2010gz} (c), with most of the data points falling
inside the yellow (shaded) error band.
In all these cases, inclusion of the NLO corrections tends to improve the
agreement.
The NLO NRQCD prediction of the $z$ distribution (d) agrees with the H1 data
in the intermediate $z$ range, but its slope appears to be somewhat too steep
at first sight.
However, the contribution due to resolved photoproduction, which is not yet
included here, is expected to fill the gap in the low-$z$ range, precisely
where it is peaked;
the overshoot of the NRQCD prediction in the upper endpoint region, which
actually turns into a breakdown at $z=1$, is an artifact of the fixed-order
treatment and may be eliminated by invoking soft collinear effective theory
\cite{Fleming:2006cd}.
We conclude from Figs.~\ref{fig:fitgraphs} and \ref{fig:other} that all
experimental data sets considered here significantly overshoot the NLO CSM
predictions, by many experimental standard deviations.
Specifically, the excess amounts to 1--2 orders of magnitude in the
case of hadroproduction [see Fig.~\ref{fig:fitgraphs}(b)] and typically a
factor of 3 in the case of photoproduction [see Fig.~\ref{fig:fitgraphs}(e)].
On the other hand, these data nicely agree with the NLO NRQCD predictions,
apart from well-understood deviations in the case of the $z$ distribution of
photoproduction [see Fig.~\ref{fig:other}(d)].
This constitutes the most rigorous evidence for the existence of CO processes
in nature and the LDME universality since the introduction of the NRQCD
factorization formalism 15 years ago \cite{Bodwin:1994jh}.

We should remark that our theoretical predictions refer to direct $J/\psi$
production, while the CDF and CMS data include prompt events and the H1 and
PHENIX data even non-prompt ones, but the resulting error is small against
our theoretical uncertainties and has no effect on our conclusions.
In fact, the fraction of $J/\psi$ events originating from the feed-down of
heavier charmonia only amounts to about 30\% \cite{Abe:1997yz} for
hadroproduction and 15\% \cite{Aaron:2010gz} for photopoprduction, and the
fraction of $J/\psi$ events from $B$ decays is negligible at HERA
\cite{Aaron:2010gz} and RHIC energies.

We thank G. Kramer for useful discussions and B. Jacak and C. Luiz da Silva for
help with the comparison to the PHENIX data \cite{Adare:2009js}.
This work was supported in part by BMBF Grant No.\ 05H09GUE, DFG Grant
No.\ KN~365/6--1, and HGF Grant No.\ HA~101.

{\it Note added.}
At the final stage of preparing this manuscript, after our results were
presented at an international conference \cite{Butenschoen:2010px}, a preprint
\cite{Ma:2010yw} appeared that also reports on a NLO calculation of $J/\psi$
hadroproduction in full NRQCD.
Adopting their inputs, we find agreement with their results for the
$^3\!S_1^{[1]}$, $^1\!S_0^{[8]}$, $^3\!S_1^{[8]}$, and $^3\!P_J^{[8]}$
contributions.

\end{document}